# A new SSO-based Algorithm for the Bi-Objective Time-constrained task Scheduling Problem in Cloud Computing Services

Chia-Ling Huang[1], Wei-Chang Yeh[2]

[1]Department of Logistics and Shipping Management, Kainan University, Taoyuan 33857, Taiwan
[2]Department of Industrial Engineering and Engineering Management, National Tsing Hua University, Hsinchu 30013, Taiwan
Corresponding author: Wei-Chang Yeh (yeh@ieee.org)

**Abstract—** Cloud computing distributes computing tasks across numerous distributed resources for large-scale calculation. The task scheduling problem is a long-standing problem in cloud-computing services with the purpose of determining the quality, availability, reliability, and ability of the cloud computing. This paper is an extension and a correction to our previous conference paper entitled "Multi Objective Scheduling in Cloud Computing Using MOSSO" published in 2018 IEEE Congress on Evolutionary Computation (CEC). More new algorithms, testing, and comparisons have been implemented to solve the bi-objective time-constrained task scheduling problem in a more efficient manner. Furthermore, this paper developed a new SSO-based algorithm called the bi-objective simplified swarm optimization (BSSO) to fix the error in previous SSO-based algorithm to address the task-scheduling problem. From the results obtained from the new experiments conducted, the proposed BSSO outperforms existing famous algorithms, e.g., NSGA-II, MOPSO, and MOSSO in the convergence, diversity, number of obtained temporary nondominated solutions, and the number of obtained real nondominated solutions. The results propound that the proposed BSSO can successfully achieve the aim of this work.

## 1. Introduction

Cloud computing is the delivery of on-demand computing resources ranging from applications to remote data centers over the internet on a pay-for-use basis.

In essence, cloud computing is a computing style in which dynamically scalable and often virtualized resources are provided as an Internet service [1]. It can be seen that the service model of cloud computing includes cloud platforms, users, applications, virtual machines, etc. In each cloud platform there are multiple platform users, each with the ability to run multiple applications in the platform. Each application corresponds to the tasks requested by a user and uses a certain quota of virtual machines, through which the application finishes and returns the tasks, thus completing the procedure.

Countless discussions and research have been conducted on the two prevailing issues in both grid and cloud computing: resource allocation and task scheduling [2-4]. The task scheduling problem revolves around exploring how the provider of cloud services assign the tasks of each client to each processor according to certain rules and regulations to ensure the cost-effectiveness of the task-scheduling process. The efficiency and performance of cloud services are usually associated with the efficiency of task scheduling, which affects not only the performance efficiency of users' tasks but the utilization efficiency of system resources as well. Hence, the interdependent relationship between the efficiency and performance of the cloud service and the efficiency of task scheduling necessitates research on the task scheduling problem of cloud computing.

The task scheduling problem of cloud-computing services assign tasks to processors to be processed. It is a NP-hard combinatorial problem, which renders it difficult to obtain the global optima within polynomial time. The greater the problem size, e.g., the number of resources or tasks, the more difficult traditional task scheduling algorithms are to solve. Hence, alongside improving traditional algorithms, many scholars have introduced machine learning algorithms, e.g., the Pareto-set cluster genetic algorithm [2] and particle swarm optimization [2], binary-code Genetic Algorithm [3] and integer-code Particle Swarm Optimization [3], simulated annealing algorithm [4], genetic algorithm [5], ant colony optimization [6], particle swarm optimization [7, 8], self-adaptive learning based particle swarm optimization [9], etc., to solve the task scheduling problem of cloud computing.

Numerous different objectives have been discussed to measure service performance, e.g., cost, reliability, makespan, power consumption [1-9]. However, when conducting research on the task assignment problem of cloud computing, most scholars emphasize on the single-objective task scheduling problem and lack the ability for these task scheduling problems with more than one objective [10-17]. By doing so, they fail to acknowledge other goals that may influence the quality of the cloud-computing service. It is thus necessary and paramount to balance various aspects when evaluating task scheduling problems [15-17]; for example, the objective of minimizing total energy consumption and the objective of minimizing makespan are a pair of conflicting objectives in real-life applications [17].

With the ever-advancing development of cloud computing,

more and more data centers have successively been established to run a large number of applications that require considerable computations and storage capacity but wastes vast amounts of energy [12, 15-17]. Reducing power consumption and cutting down energy cost has become a primary concern for today's data center operators. Thus, striking a balance between reducing energy consumption and maintaining high computation capacity has become a timely and important challenge that calls for concern.

Furthermore, with more media and public attention shifting onto progressively severe environmental issues, governments worldwide have now adopted a stronger environmental protection stance [12, 15-17]. This puts increasing pressure on enterprises to pursue higher output and also focus on minimizing power consumption [12, 15-17].

Stemming from real-life concerns as mentioned above, our previous work considered a bi-objective time-constrained task scheduling problem to measure the performance of a certain task schedule plan with two objectives: the quality of the cloud computing service, which is in terms of the makespan and the environmental problem based on the energy consumption [12, 15-17].

There are two different types of algorithms used to solve multi-objective problems. While one converts multi-objective problems to ones that are single objective in nature through methods, e.g., as ε-constraint, LP-metrics, goal programming, etc., the other solves multi-objective problems based on the concept of Pareto optimality. The latter is the one we adapted here and in our previous study [10-17].

Moreover, a new algorithm called the multi-objective simplified swarm optimization (MOSSO) is proposed to solve the above problem [17]. However, there are some errors in the MOSSO source code and the performance comparison is limited to MOSSO and MOPSO despite both having an iterative local search [17].

In order to rectify our previous source code with new concepts based on the Pareto optimality to solve the bi-objective time-constrained task scheduling, we proposed a new algorithm called the bi-objective simplified swarm optimization (BSSO). It draws from the SSO update mechanism to generate offspring [18], the crowding distance to rank nondominated solutions [19], the new hybrid elite selection to select parents [20], and the limited number of nondominated solutions adapted from multi-objective particle swarm optimization (MOPSO) serves to guide the update [13].

The remainder of this paper is organized as follows. Section 2 presents notations, assumptions, and the mathematical modeling of the proposed bi-objective time-constrained task scheduling problem to address energy consumption and service quality which is in terms of the makespan. Section 3 introduces the simplified swarm optimization (SSO) [18], the concept of the crowd distance [19], and the traditional elite selection [20]. The proposed BSSO is presented in Section 4 together with the discussion of its novelties and pseudo code. Section 5 compares the proposed BSSO in nine different parameter settings with the MOSSO proposed in [17], the MOPSO proposed in [13], and the famous NSGA-II [19] in three benchmark problems with small-size, medium-size and large size from the aspect of the convergence, diversity, number of obtained nondominated solutions to demonstrate the performance of the BSSO. Our conclusion is given in Section 6.

## 2. Notations, Assumptions and Mathematical Problem Description

The formal mathematical model of the bi-objective constrained task scheduling problem in cloud-computing services [12, 15-17] is described as follows, together with notations and assumptions used in the problem and algorithms.

### 2.1. Notations

The following notations are used:

- $|\bullet|$ : number of elements in $\bullet$.
- $N_{var}$ : number of tasks used in the test problem.
- $N_{cpu}$ : the number of processors contained in the given data center;
- $N_{run}$ : number of runs for the algorithms.
- $N_{gen}$ : number of generations in each run.
- $N_{sol}$ : number of solutions in each generation.
- $N_{non}$ : number of selected temporary nondominated solutions.
- $X_i$ : $i$th solution.
- $x_{i,j}$ : $j$th variable in $X_i$.
- $P_i$ : the best solution among all solutions updated based on $X_i$ in SSO.
- $p_{i,j}$ : $j$th variable in $P_i$.
- $gBest$ : index of the best solution among all solutions in SSO, i.e., $F(P_{gBest})$ is better than or equal to $F(P_i)$ for $i = 1, 2, …, N_{sol}$.
- $\rho_I$ : random number generated uniformly within interval $I$.
- $c_g, c_p, c_w, c_r$ : positive parameters used in SSO with $c_g + c_p + c_w + c_r = 1$.
- $C_g, C_p, C_w$ : $C_g = c_g$, $C_p = C_g + c_p$, and $C_w = C_p + c_w$.
- $F_l(\bullet)$ : $l$th fitness function value of solution $\bullet$.
- $Max(\bullet)$ : maximal value of $\bullet$, i.e., $Max(F_l)$ is the maximal value of the $l$th objective function.
- $Min(\bullet)$ : minimal value of $\bullet$, i.e., $Min(F_l)$ is the minimal value of the $l$th objective function.
- $S_t$ : set of selected solutions from $\Pi_t$ to generate new solutions in the $(t+1)^{th}$ generation. Note that $S_1 = \Pi_1$ and $|S_t| = N_{sol}$ for $i = 1, 2, …, N_{gen}$.
- $size_i$ : size of the task $i$ for $i = 1, 2, ..., N_{var}$
- $start_i$ : start time of the task $i$ for $i = 1, 2, ..., N_{var}$
- $speed_j$ : execution speed of the processor $j$ for $j = 1, 2, ..., N_{cpu}$
- $e_j$ : energy consumption per unit time of the processor $j$ for $j = 1, 2, ..., N_{cpu}$;
- $T_{ub}$ : deadline constraint of task scheduling
- $t_{i,j}$ : processing time $t_{i,j} =$

$$t_{i,j} = \begin{cases} size_i / speed_j & \text{if task } i \text{ is processed on processor } j \\ 0 & \text{otherwise} \end{cases}$$

for $i = 1, 2, ..., N_{var}$ and $j = 1, 2, ..., N_{cpu}$

### 2.2. Assumptions

In cloud-computing services, all requests from cloud-computing users are collected as jobs and subsequently divided into multiple tasks which are then executed by data centers with multiple processors [1-12, 15-17]. To simplify our model without foregoing generality, the following assumptions are utilized to prompt the bi-objective constrained task scheduling problem to focus on the energy consumption and makespan of the cloud-computing service [1-12, 15-17].
1. All tasks are available, independent, and equal in importance with two attributes $size_i$ and $time_i$ for $i = 1, 2, ..., N_{var}$ for processing simultaneously at time zero.
2. Each processor is available at any time with two attributes, denoted as $speed_j$ and $power_j$, $j = 1, 2, ..., N_{cpu}$, and cannot process two or more tasks simultaneously.
3. All processing times include set-up times.
4. Task pre-emption and the splitting of tasks are not permitted.
5. There is an infinite buffer between the processors.

### 2.3. The mathematical model

A feasible task scheduling solution is a plan that assigns tasks to processors in a sequence such that the makespan, which is the time at which the final task is complete, is less than a predefined limited. Let solution $X = (x_1, x_2, ..., x_{Nvar})$ be a feasible task scheduling plan and $x_i$ be the type of processor assigned to the task $i$ for $i = 1, 2, ..., N_{var}$.

To maintain efficiency and ensure environmental protection, the bi-objective constrained task scheduling problem considered here is to assign $N_{var}$ tasks to $N_{cpu}$ processors to minimize both the energy consumption and makespan so that the makespan is not overdue: [17]

$$\text{Min } F_e(X) = \sum_{i=1}^{n}(t_{i,x_i} \cdot e_{x_i}) \tag{1}$$

$$\text{Min } F_m(X) = \max_j \sum_{i=1}^{n} t_{i,j}. \tag{2}$$

$$\text{s.t. } F_m(X) \leq T_{ub}. \tag{3}$$

Eq. (1) is the total energy consumption of a task scheduling plan and it is the sum of the usage of energy of all tasks assigned to processors based on $X$. Note that the energy consumption of each task is the product of the power cost per unit time and the running time of which is equal to size of the task divided by the execution speed, as shown in formula (4).

$$t_{i,j} = \begin{cases} size_i / speed_j & \text{if task } i \text{ is processed on processor } j \\ 0 & \text{otherwise} \end{cases} \tag{4}$$

Eq. (2) is the makespan which is the time of the last task is finished. Eq. (3) is a time constraint for each job such that each job must be finished before the deadline $T_{ub}$.

## 3. SSO, Crowding Distance, and Elite Selection

The proposed BSSO is based on SSO [18] to update solutions from generation to generation intelligently, the crowding distance [19] to rank temporary nondominated solutions systematically, and the elite selection [20] to select solutions to act as parents. Hence, SSO, crowding distance, and elite selection are introduced briefly in this section.

### 3.1. Simplified Swarm Optimization

SSO developed by Yeh in 2009 [18] is a simple but powerful machine leaning algorithm hybrid of a-leader-solution swarm intelligence and the population-based evolutionary computation.

In traditional SSO, all variables need to be updated (called the all-variable update in SSO) such that the $j$th variable of the $i$th solution (i.e., $x_{i,j}$) is obtained from either the $j$th variable of the $P_{gBest}$ (i.e., $p_{gBest,j}$) with a probability $c_g$, of the $P_i$ with a probability $c_p$, of its current value (i.e., $x_{i,j}$) with a probability $c_w$, or a randomly generated feasible new value (say $x$) with a probability $c_r$, where the $P_{gBest}$ is the best solution among all existing solutions, the $P_i$ is the best $i$th solution in its evolutionary history, and $c_g + c_p + c_w + c_r = 1$.

The above update mechanism of SSO is very simple, efficient, flexible [17, 18, 21-26], and can be presented as a stepwise-function update:

$$x_{i,j} = \begin{cases} p_{gBest,j} & \text{if } \rho_{[0,1]} \in [0, C_g) \\ p_{i,j} & \text{if } \rho_{[0,1]} \in [C_g, C_p) \\ x_{i,j} & \text{if } \rho_{[0,1]} \in [C_p, C_w) \\ x & \text{if } \rho_{[0,1]} \in [C_w, 1] \end{cases} \tag{5}$$

where $C_g = c_g$, $C_p = C_g + c_p$, and $C_w = C_p + c_w$.

The SSO pseudo code is provided below [17, 18, 21-25]:

**STEP S0.** Generate $X_i$ randomly, find $gBest$, and let $t=1$, $k=1$, and $P_i = X_i$ for $i = 1, 2, ..., N_{sol}$.
**STEP S1.** Update $X_k$.
**STEP S2.** If $F(X_k)$ is better than $F(P_k)$, let $P_k = X_k$. Otherwise, go to STEP S5.
**STEP S3.** If $F(P_k)$ is better than $F(P_{gBest})$, let $gBest = k$.
**STEP S4.** If $k < N_{sol}$, let $k = k+1$ and go to STEP S1.
**STEP S5.** If $t < N_{gen}$, let $t = t+1$, $k=1$, and go to STEP S1. Otherwise, halt.

The stepwise-function update mechanism is very simple and efficient, but it is also very powerful and with various success applications, e.g., the redundancy allocation problem [21, 22], the disassembly sequencing problem [23, 24], artificial neural network [25], data mining [26], energy problems [27, 28], etc. Moreover, the stepwise-function update mechanism is easier to customize by either replacing any item of its stepwise function with other algorithms [22, 25], even hybrid algorithms [27] in sequence or in parallel [28], to solve various problems than to customize other algorithms [17, 18, 21-28].

### 3.2. Crowding Distance

Let

$$\hat{d}_{l,i} = Min\{\frac{F_l(X_i) - F_l(X_j)}{Max(F_l) - Min(F_l)}\} \quad (6)$$

be the shortest normalized Euclidean distance between the $i^{th}$ temporary nondominated solution and any other temporary nondominated solutions based on the $l^{th}$ objective function, where $Min(F_l)$ is the minimal value of the $l$th objective function, and $Max(F_l)$ is the maximal value of the $l$th objective function.

The crowding distance is the sum of the individual distance and can be calculated as follows [19]:

$$\sum_i \sqrt{\hat{d}_{1,i}^2 + \hat{d}_{2,i}^2}, \quad (7)$$

for all temporary nondominated solutions $X_i$. The crowding distance is only used to rank temporary nondominated solutions for selecting as parents if the total number of determine are over a limited value which is defined as $N_{non}$ and $N_{non} = N_{sol}$ in this study.

### 3.3. Notations The Elite Selection

Among numerous different selection policies, the elite selection is the simplest and thus adopted in the proposed BSSO. The elite selection chooses best solutions in the current generation to generate and update solutions for the next generation [20]. For example, let $N_{sol} = 50$, $X_1, X_2, ..., X_{100}$ be solutions needed to select. Elite selection ranks and selects the best 500 solutions among $X_1, X_2, ..., X_{100}$ and renumbers these selected solutions as $X_1, X_2, ..., X_{50}$.

In the BSSO, these best solutions are temporary nondominated solutions. Due to the number of temporary nondominated solutions may be less than or larger than the number of parents, a new elite selection called the hybrid elite selection is developed and used in the proposed BSSO and the details is discussed in Section 4.3.

## 4. Proposed Algorithm

The proposed BSSO is a population-based, all-variable update, and stepwise function-based method, i.e., there are a fixed number of solutions in each generation and all variables must be updated based on the stepwise function for each solution. Details of the proposed BSSO are discussed in this section.

### 4.1. Solution Structure

The first step of most machine learning algorithms is to define the solution structure [1-9, 21-28]. A solution in the BSSO for the proposed problem is defined as a vector, and the value of the $i$th coordinate is the processor utilized to process the $i$th task. For example, let $X_5 = (3, 4, 6, 4)$ be the $5^{th}$ solution in the $10^{th}$ generation of the problem with 4 tasks. In $X_5$, tasks 1, 2, 3, and 4 are processed by processors 3, 4, 6, and 4.

### 4.2. Novel Update Mechanism

The second step in developing machine learning algorithms is to create an update mechanism to update solutions [1-9, 21-28]. The stepwise update function is a unique update mechanism of SSO [17, 18, 21-28]. In the single objective problem, there is only one *gBest* for the traditional SSO. However, in the multi-objective problem the number of temporary nondominated solutions is not always only one, and some temporary nondominated solutions may not be temporary nondominated solutions in the next generation [17]. Note that a nondominated solution is not dominated by any other solution, while a temporary nondominated solution is a solution nondominated by other found solutions that are temporary in the current generation and it may be dominated by other updated solutions later [13, 17, 19].

Hence, in the proposed BSSO the role of *gBest* is removed completely and the *pBest* used in the BSSO is not the original definition in SSO. The *pBest* for each solution in the proposed BSSO is selected from temporary nondominated solutions, i.e., there is no need to follow its previous best predecessor. The stepwise update function used in the proposed bSSO is listed below for multi-objective problems for each solution $X_i$, with $i = 1, 2, ..., N_{sol}$:

$$x_{i,j} = \begin{cases} x_i^* & \text{if } \rho_{[0,1]} \in [0, C_p) \\ x_{i,j} & \text{if } \rho_{[0,1]} \in [C_p, C_w) \\ x & \text{otherwise} \end{cases} \quad (8)$$

where $X^* = (x_1^*, x_2^*, ..., x_k^*)$ is one of the temporary nondominated solutions selected randomly, $\rho_{[0,1]}$ is a random number generated uniformly in [0, 1], $x$ is a random integer generated from 0, 1, 2,..., $N_{var}$.

For example, let $X_6 = (1, 2, 3, 2, 4)$ and $X^* = (2, 1, 4, 3, 3)$ be a temporary nondominated solution selected randomly. Let $C_p = 0.50$, $C_w = 0.95$, and $\rho = (\rho_1, \rho_2, \rho_3, \rho_4, \rho_5) = (0.32, 0.75, 0.47, 0.99, 0.23)$. The procedure to update $X_6$ based on the stepwise function provided in Eq. (8) is demonstrated in Table 1.

**Table 1.** Example of the update process in the proposed BSSO.

| Variable | 1 | 2 | 3 | 4 | 5 |
|---|---|---|---|---|---|
| $X_5$ | 1 | **2** | 3 | 2 | 4 |
| $X^*$ | **2** | 1 | **4** | 3 | **3** |
| ρ | 0.32 | 0.75 | 0.47 | 0.99 | 0.23 |
| New $X_5$ | 2 | 2 | 4 | 4# | 3 |

"#" indicates that the corresponding feasible value is generated randomly

From the above example, the simplicity, convenience, and efficiency of the SSO can also be found in the update mechanism of the proposed BSSO.

### 4.3. Hybrid Elite Selection

The last step is to determine the selection policy to decide which solutions, i.e., parents, are selected to generate solutions in the next generation. In the proposed BSSO, the hybrid selection is harnessed.

Let $\pi_t$ be a set to store the temporary nondominated solutions found in the $t^{th}$ generation. It is impossible to have all temporary nondominated solutions since its number is infinite and also as temporary nondominated solutions may not be real nondominated solutions. The value of $|\pi_t|$ is limited to $N_{sol}$ and parts of temporary nondominated solutions are abandoned to keep $|\pi_t| = N_{sol}$.

If $N_{sol} \leq |\pi_t|$, the crowding distances need to be calculated for each temporary nondominated solution and only the best $N_{sol}$ solutions will be selected from $\pi_t$ to be parents. However, all temporary nondominated solutions in $\pi_t$ are used to serve as parents and ($N_{sol} - |\pi_t|$) solutions are selected randomly from offspring if $N_{sol} > |\pi_t|$. As discussed above, there are always $N_{sol}$ parents for each generation and these temporary nondominated solutions are usually chosen first to serve their role as parents.

Note that these temporary nondominated solutions are abandoned if they are not selected as parents.

### 4.4. Group Comparison

From Section 4.3, the temporary nondominated solutions play a paramount role in most multi-objective algorithms. Up to now, the most popular method to achieve the above goal is the pairwise comparison, which takes $O(N^2)$ [13, 17, 19] for each solution in each generation where N is the number of solutions from which we determine temporary nondominated solutions. Hence, the corresponding related time complexity of the BSSO and NSGA-II are both $O(4N_{sol}^2)$. Hence, the most time-consuming aspect of solving multi-objective problems using machine learning algorithms is the search for all temporary nondominated solutions from all offspring.

In the proposed BSSO, the parents are selected from temporary nondominated solutions found in the previous generation and offspring generated in the current generation. To reduce the computation burden, a new method called the group comparison is proposed in BSSO. All temporary nondominated solutions in offspring are found first using the pairwise comparison, which takes $O(N_{sol}^2)$ as the number of offspring is $N_{sol}$. Then, the temporary nondominated solutions obtained in the previous generations are compared with the new temporary nondominated solutions found in the current generation. The number of both sets of temporary nondominated solutions are at most $N_{sol}$, i.e., the time complexity is $O(N_{sol} \log(N_{sol}))$ based on the merge sort.

Hence, the time complexity is $O(N_{sol} \log(N_{sol})) + O(N_{sol}^2) = O(N_{sol}^2)$ which is only one quarter of the pairwise comparison

### 4.5. Proposed BSSO

The procedures of the proposed BSSO based on the solution structure, update mechanism, hybrid elite selection, and group comparison discussed in this section are presented in pseudo code as follows.

**PROCEDURE BSSO**
**STEP 0.** Create initial population $X_i$ randomly for $i = 1, 2, \ldots, N_{sol}$ and let $t=2$.
**STEP 1.** Let $\pi_t$ be all temporary nondominated solutions in $S^* = \{X_k \mid$ for $k = 1, 2, \ldots, 2N_{sol}\}$ and $X_{N_{sol}+k}$ is updated from $X_k$ based on Eq. (8) for $k = 1, 2, \ldots, N_{sol}$.
**STEP 2.** If $N_{sol} \leq |\pi_t|$, let $S = \{$ top $N_{sol}$ solutions in $\pi_t$ based on crowding distances$\}$ and go to STEP 4.
**STEP 3.** Let $S = \pi_t \cup \{ N_{sol} - |\pi_t| $ solutions selected randomly from $S^* - \pi_t\}$.
**STEP 4.** Re-index these solutions in $S$ such that $S = \{X_k \mid$ for $k = 1, 2, \ldots, N_{sol}\}$.
**STEP 5.** If $t < N_{gen}$, then let $t = t+1$ and go back to STEP 1. Otherwise, halt.

## 5. Simulation results and discussion

Numerous experiments of the parameter-setting procedure and the performances of BSSO were carried out on nine different settings for the three different-size benchmark problems used in [17]. The experimental results were compared with those obtained using MOPSO [13], MOSSO [17], and NSGA-II [19]. Note that NSGA-II is the current best algorithm for multi-objective problems.

### 5.1. Parameter-Settings

All machine learning algorithms have parameters in their update procedures and/or the selection procedure. Thus, there is a need to tune parameters for better results. To determine the best parameters $c_p$ and $c_w$ of the proposed BSSO, three different levels of two factors $c_p$ and $c_w$: the low value 0.1, the middle value 0.3, and the high value 0.5 have been combined, e.g., nine different parameter settings as shown in Table 2.

**Table 2.** Paratter setting of the proposed BSSO.

| Alg. | 0 | 1 | 2 | 3 | 4 | 5 | 6 | 7 | 8 |
|---|---|---|---|---|---|---|---|---|---|
| $C_p$ | 0.1 | 0.1 | 0.1 | 0.3 | 0.3 | 0.3 | 0.5 | 0.5 | 0.5 |
| $C_w$ | 0.2 | 0.4 | 0.6 | 0.4 | 0.6 | 0.8 | 0.6 | 0.8 | 1.0 |

Note that $C_p = c_p$ and $C_w = c_p + c_w$. The following provided the parameter-settings for the other algorithms:
MOSSO (Alg. 9): $C_g = 0.1 + 0.3t/N_{gen}$, $C_p = 0.3 + 0.4t/N_{gen}$, $C_w = 0.4 + 0.5t/N_{gen}$, where $t$ is the current generation number [17].
MOPSO (Alg. 10): $w = 0.871111$, $c_1 = 1.496180$, $c_2 = 1.496180$ [17]
NSGA-II (Alg. 11): $c_{crossover} = 0.7$, $c_{mutation} = 0.3$

### 5.2. Experimental Environments

To demonstrate the performance of the proposed BSSO and select the best parameter settings, the BSSO with nine parameter settings were utilized for three task scheduling benchmarks [17], namely ($N_{task}$, $N_{cpu}$) = (20, 5), (50, 10), and (100, 20) and the deadlines were all set 30 for each test [17].

The following lists the special characteristics of the processor speeds (in MIPs), the energy consumptions (in KW/per unit time) and the task sizes (in MI) in these three benchmark problems [17]:
1. Each processor speed is generated between 1000 and 10000 (MIPs) randomly and the largest speed is ten times the smallest one.
2. The power consumptions grow polynomial as the speed of processors grow and the value range is between 0.3 and 30 (KW) per unit time.
3. The values of task sizes are between 5000-15000(MIs).

Alongside the proposed BSSO with nine parameter settings (Alg. 0-8), there are three other multi-objective algorithms: MOSSO (Alg. 9) [17], MOPSO (Alg. 10) [13, 17], and NSGA-II (Alg. 11) [19] were tested and compared to further validate the superiority of the proposed BSSO. The NSGA-II [19] is the current best multi-objective algorithm and is based on genetic algorithm while the MOPSO is based on particle swarm optimization [13], and MOSSO on SSO [17].

The proposed BSSO with its nine parameter settings and the NSGA-II have no iterative local search to improve the solution quality. To ensure a fair comparison, the iterative local search is removed from both MOPSO and MOSSO. All algorithms are coded in DEV C++ on a 64-bit Windows 10 PC, implemented on an Intel Core i7-6650U CPU @ 2.20 GHz notebook with 16 GB of memory.

Also, for a fair comparison between all algorithms, $N_{sol}$ = $N_{non}$ = 50, $N_{gen}$ = 1000, and $N_{run}$ = 500, i.e., the same solution number, generation number, size of external repository, and run number for each benchmark problem respectively were used. Furthermore, the calculation number of the fitness function of all algorithms was limited to $N_{sol} \times N_{gen}$ = 50,000, 100,000, and 150,000 in each run for the small-size, medium-size, and large-size benchmarks, respectively.

Noe that the reason for the large value of $N_{run}$ is to simulate the Pareto front and the details are discussed in Section 5.3.

### 5.3. Performance metrics

The convergence metrics and diversity metrics both are always used to evaluate the performances of the multi-objective algorithms in the solution quality. Among these metrics, the general distance (GD) introduced by Van Veldhuizen et al. [29] and spacing (SP) introduced by Schott [30] are two general indexes for the convergence metrics and diversity metrics, respectively. Let $d_i$ be the shortest Euclidean distance between the $i^{th}$ temporary nondominated solution and the Pareto front, and $\underline{d}$ be the average sum of all $d_i$ for all $i$. The GD and SP are defined below:

$$GD = \frac{\sqrt{\sum_{i=1}^{N_{sol}} d_i^2}}{N_{sol}}, \quad (9)$$

$$SP = \sqrt{\frac{\sum_{i=1}^{N_{sol}} (\underline{d} - d_i)^2}{N_{sol} - 1}}. \quad (10)$$

The GD is the average of sum of the squares of $d_i$ and the SP is very similar to the standard deviation in probability theory and statistics. If all temporary nondominated solutions are real nondominated solutions, we have GD=0. The solutions are equality far from $\underline{d}$, we have SP=0. Hence, in general, the smaller the SP is, the higher the diversity of solutions along the Pareto front and the better the solution quality becomes [13, 17, 19, 29, 30].

The Pareto front is needed for both the GD and SP in calculating from their formulas [29, 30]. Unfortunately, it requires infinite nondominated solutions to form the Pareto front and this is impossible to accomplish even with exhaustive methods for the task scheduling problem of cloud computing, which is a NP-hard problem [12, 15-17]. Rather than a real Pareto Front, a simulated Pareto Front is implemented by collecting all temporary nondominated solutions in the final generation from all different algorithm with different values of $N_{sol}$ for the same problem.

There are 12 algorithms with three different $N_{sol}$ =50, 100, and 150 for $N_{run}$=500, i.e., 12 × 500 × (50 + 100 + 150) = 1,800,000 solutions obtained in the end for each test problem. All temporary nondominated solutions are found from these 1,800,000 final solutions to create a simulated Pareto Front to calculate the GD and SP.

### 5.4. Numerical Results

All numerical results attained in the experiments are listed in this section. Tables 3-5 list the averages and standard deviations of obtained number of temporary nondominated solutions (denoted by $N_n$), the obtained number of nondominated solutions in the Pareto front (denoted by $N_p$), the converge metric GD, and the diversity metric SP, the required run time (denoted by T), the energy consumption values (denoted by $F_e$), and the makspan values (denoted by $F_m$) for three different size benchmark problems with $N_{sol}$=50, 100, and 150, respectively. In these tables, the best of all algorithms is indicated in bold.

From Tables 3-5, we have the following general observations:
1. The lower $c_r$ value, the better performance. In the small-size problem, Alg. 7, i.e., the proposed BSSO with $c_p$=0.5 and $c_w$=0.3, is the best among all 12 algorithms. However, Alg. 8, i.e., the proposed BSSO with $c_p$=0.5 and $c_w$=0.5, is the best one for the medium and large size problems. The reason is that the number of real nondominated solutions is infinite. Even though in the update mechanism of the proposed BSSO when $c_r$=0 there is only an exchange of information between the current solution itself and one of selected temporary nondominated solutions, it is already able to update the current solution to a better solution without needing any random movement.

2. The larger size of the problem, e.g., $N_{task}$, the fewer number of obtained nondominated solutions, e.g., $N_n$ and $N_p$. There are two reasons why this is the case: 1). due to the characteristic of the NP-hard problems, i. e., the larger size of the NP-hard problem, the more difficult it is to solve; 2) it is more difficult to find nondominated solutions for larger problems with the same deadline of 30 for all problems.

3. The larger $N_{sol}$, the more likely it is to find more nondominated solutions, i.e., the larger $N_n$ and $N_p$, for the best algorithm among these algorithms no matter the size of the problem. Hence, it is an effective method to have larger value of $N_{sol}$ if we intend to find more nondominated solutions.

4. $N_{task}$, the fewer number of obtained nondominated solutions, e.g.,. There are two reasons of the above: 1). due to the characteristic of the NP-hard problems, i. e., the larger size of the NP-hard problem, the more difficult to solve; 2) it is more difficult to find nondominated solutions for a larger problems with the same deadline which is set 30 for all problems.

5. The smaller value of $N_p$, the shorter run time. The most time consuming aspect of finding nondominated solutions is filtering out these nondominated solutions from current solutions. Hence, the new method called the group comparison is proposed in this study to find these nondominated solutions from current solutions. However, even the proposed group comparison is more efficient than the traditional pairwise comparison on average, it still needs $O(N_p^2)$ to achieve the goal.

6. There is no special pattern in solution qualities, e.g., the value of GD, SP, $N_n$, and $N_p$, from the final average valuers of the energy consumption and the makespan.

7. The one with the better number of obtained nondominated solutions also has better DP and SP values.

8. The MOPSO (Alg. 10) [13] and the original MOSSO (Alg. 9) [17] share one common factor: each solution must inherit and update based on its predecessor (parent) and its *pBest* and this is the main reason that it is less likely to find new nondominated solutions. The above observation is coincident to that observed in item 1. Hence, the proposed BSSO and the NSGA-II (Alg. 11) [19] are much better than MOSSO (Alg. 9) [17] and MOPSO (Alg. 10) [13] in solution quality.

In general, the proposed BSSO without $c_r$ has a more satisfying performance in all aspect of measures.

**TABLE 3.** Result for small size problem.

| $N_{sol}$ | Alg | Avg($N_n$) | Std($N_n$) | Avg($N_p$) | Std($N_p$) | Avg(GD) | Std(GD) | Avg(SP) | Std(SP) | Avg(T) | Std(T) | Avg($F_e$) | Std($F_e$) | Avg($F_m$) | Std($F_m$) |
|---|---|---|---|---|---|---|---|---|---|---|---|---|---|---|---|
| 50 | 0 | 48.34 | 2.22 | 0.012 | 0.109 | 0.565 | 0.611 | 3.527 | 4.343 | 2.8954 | 0.5839 | 11937.28 | 261.3206 | 2539.957 | 4798.1386 |
| | 1 | 49.634 | 0.904 | 0.018 | 0.133 | 0.26 | 0.25 | 1.475 | 1.781 | 3.4103 | 0.6708 | 11689.93 | 272.5046 | 1015.79 | 1926.8740 |
| | 2 | 49.966 | 0.202 | 0.058 | 0.234 | 0.138 | **0.059** | 0.722 | **0.445** | 4.1789 | 0.6901 | 11185.89 | 273.4064 | 838.2936 | 632.7848 |
| | 3 | 49.494 | 1.165 | 0.012 | 0.109 | 0.311 | 0.349 | 1.819 | 2.495 | 3.2951 | 0.6579 | 11766.6 | 269.763 | 1208.26 | 2270.2913 |
| | 4 | 49.936 | 0.276 | 0.034 | 0.202 | 0.17 | 0.073 | 0.915 | 0.547 | 3.9837 | 0.6919 | 11317.03 | 289.0288 | **779.7255** | 448.9989 |
| | 5 | **50** | **0** | **0.284** | 0.544 | 0.118 | 0.065 | **0.71** | 0.478 | 4.7604 | 0.7535 | 10603.56 | 229.703 | 938.6334 | **22.5991** |
| | 6 | 49.968 | 0.208 | 0.054 | 0.235 | 0.153 | 0.101 | 0.821 | 0.734 | 4.2185 | 0.6928 | 11230.14 | 290.7808 | 817.7619 | 631.6859 |
| | 7 | **50** | **0** | 0.218 | 0.464 | 0.121 | 0.064 | 0.722 | 0.475 | **4.86** | 0.7531 | 10561.73 | 236.0619 | 936.6982 | 24.2198 |
| | 8 | 49.7 | 0.766 | 0.188 | 0.457 | 0.153 | 0.103 | 0.776 | 0.598 | 4.4077 | 0.8044 | 10088.1 | 550.3918 | 1010.433 | 888.7098 |
| | 9 | 9.274 | 2.026 | 0 | 0 | 2.947 | 0.908 | 17.366 | 6.314 | 3.1238 | 0.5445 | 12176.37 | 230.151 | 29378.78 | 16620.2163 |
| | 10 | 1.23 | 0.508 | 0 | 0 | 9.914 | 0.122 | 8.417 | 4.978 | 1.2282 | 0.187 | **9833.038** | **24.8444** | 488648.8 | 10786.0609 |
| | 11 | 17.22 | 3.216 | 0 | 0 | 2.213 | 1.27 | 13.107 | 8.676 | 0.0102 | **0.0202** | 11963 | 487.9212 | 19913.23 | 20625.1095 |
| 100 | 0 | 61.856 | 5.503 | 0.024 | 0.153 | 1.996 | 0.502 | 18.548 | 4.721 | 10.3052 | 1.4032 | 24162.67 | 365.2535 | 48016.98 | 20979.7233 |
| | 1 | 71.016 | 5.808 | 0.05 | 0.227 | 1.284 | 0.506 | 12.06 | 4.968 | 10.2902 | 1.3693 | 23633.28 | 438.6975 | 24077.66 | 15460.6944 |
| | 2 | 95.07 | 4.779 | 0.184 | 0.463 | 0.289 | 0.306 | 2.712 | 3.064 | 11.282 | 1.6767 | 22010.09 | 475.5156 | 4596.549 | 5938.6783 |
| | 3 | 71.316 | 5.619 | 0.064 | 0.253 | 1.347 | 0.508 | 12.678 | 4.946 | 10.5688 | 1.3536 | 23621.2 | 426.826 | 26436.17 | 16447.3743 |
| | 4 | 88.894 | 6.061 | 0.116 | 0.362 | 0.537 | 0.387 | 5.102 | 3.839 | 10.8808 | 1.4016 | 22401.5 | 480.8059 | 8531.313 | 8913.8709 |
| | 5 | 99.98 | 0.165 | 0.668 | 0.824 | 0.061 | 0.037 | 0.514 | 0.382 | 21.7762 | 2.6418 | 19982.09 | 420.3826 | 1891.765 | 452.4866 |
| | 6 | 98.638 | 2.135 | 0.278 | 0.549 | 0.137 | 0.174 | 1.212 | 1.755 | 13.335 | 2.399 | 21646.2 | 473.8394 | 2588.289 | 3570.9857 |
| | 7 | **99.996** | **0.088** | 0.714 | 0.849 | 0.062 | **0.035** | 0.521 | **0.369** | 23.0932 | 2.6965 | 19860.81 | 428.8591 | **1879.873** | **57.8411** |
| | 8 | 99.89 | 0.546 | **1.354** | 1.173 | **0.059** | 0.04 | **0.497** | 0.401 | 26.0226 | 4.3313 | **19086.17** | 614.0985 | 2056.674 | 772.1143 |
| | 9 | 12.042 | 2.554 | 0 | 0 | 2.113 | 0.454 | 17.758 | 4.29 | 12.0856 | 1.6538 | 24347.86 | 307.4949 | 57505.36 | 23415.3032 |
| | 10 | 1.58 | 0.699 | 0 | 0 | 7.01 | 0.055 | 9.271 | 3.117 | 4.7402 | 0.4797 | 19666.68 | **33.48** | 977197.9 | 13878.8233 |
| | 11 | 21.096 | 3.25 | 0 | 0 | 2.25 | 0.777 | 19.435 | 7.076 | 0.0392 | **0.0489** | 24289.14 | 713.0953 | 65231.95 | 40247.3188 |
| 150 | 0 | 67.96 | 5.77 | 0.036 | 0.197 | 1.97 | 0.343 | 21.901 | 3.817 | 23.0904 | 3.509 | 36497.8 | 446.2135 | 99073.14 | 31223.9014 |
| | 1 | 76.334 | 6.347 | 0.078 | 0.276 | 1.552 | 0.349 | 17.458 | 4.014 | 22.8375 | 3.3043 | 35809.46 | 542.9496 | 69700.06 | 27524.2792 |
| | 2 | 102.624 | 7.73 | 0.22 | 0.473 | 0.929 | 0.312 | 10.699 | 3.723 | 23.8908 | 3.3867 | 33666.71 | 657.8649 | 36715.2 | 20668.0023 |
| | 3 | 79.2 | 5.726 | 0.098 | 0.304 | 1.465 | 0.327 | 16.518 | 3.814 | 23.6139 | 3.4082 | 35679.86 | 536.1647 | 64560.6 | 24571.6138 |
| | 4 | 98.296 | 6.9 | 0.166 | 0.413 | 0.932 | 0.276 | 10.655 | 3.313 | 24.0771 | 3.3979 | 33969.23 | 669.9933 | 36254.89 | 18845.1508 |
| | 5 | 149.92 | 0.427 | 1.362 | 1.177 | 0.042 | **0.028** | 0.441 | 0.361 | 43.0185 | 7.8536 | 29176.09 | 593.4306 | **2833.142** | 627.5616 |
| | 6 | 122.092 | 7.632 | 0.408 | 0.631 | 0.591 | 0.254 | 6.907 | 3.087 | 25.971 | 3.4265 | 32432.3 | 657.7752 | 20747.74 | 14131.2079 |
| | 7 | **149.954** | **0.277** | 1.47 | 1.151 | 0.04 | **0.028** | 0.414 | 0.356 | 51.0654 | 7.7838 | 28979.39 | 598.0727 | 2870.726 | 638.1370 |
| | 8 | 149.904 | 0.602 | **3.548** | 2.013 | **0.036** | **0.028** | **0.39** | **0.347** | 74.6991 | 13.5849 | **27451.57** | 612.008 | 3233.779 | 995.0084 |
| | 9 | 13.692 | 2.496 | 0 | 0 | 1.744 | 0.289 | 18.011 | 3.32 | 26.7054 | 4.1966 | 36486.46 | 388.5253 | 87266.1 | 26859.1298 |
| | 10 | 1.986 | 0.921 | 0 | 0 | 5.724 | 0.038 | 9.331 | 2.551 | 10.4472 | 1.2398 | 29494.15 | **43.4111** | 1466864 | 18034.0917 |
| | 11 | 23.45 | 3.732 | 0 | 0 | 2.109 | 0.579 | 22.251 | 6.213 | 0.0855 | **0.0743** | 36643.47 | 992.1199 | 120913.5 | 60590.6631 |

**TABLE 4.** Result for medium size problem.

| $N_{sol}$ | Alg | Avg($N_n$) | Std($N_n$) | Avg($N_p$) | Std($N_p$) | Avg(GD) | Std(GD) | Avg(SP) | Std(SP) | Avg(T) | Std(T) | Avg($F_e$) | Std($F_e$) | Avg($F_m$) | Std($F_m$) |
|---|---|---|---|---|---|---|---|---|---|---|---|---|---|---|---|
| 50 | 0 | 24.472 | 4.177 | 0 | 0 | 20.001 | 2.216 | 109.497 | 7.657 | 4.5751 | 0.7871 | 32537.48 | 456.8619 | 196678.1 | 37821.6436 |
| | 1 | 25.954 | 4.105 | 0 | 0 | 18.074 | 2.375 | 103.29 | 9.431 | 4.403 | 0.7533 | 32405.09 | 566.3836 | 167400.6 | 37900.9380 |
| | 2 | 29.672 | 4.648 | 0 | 0 | 14.629 | 2.778 | 89.001 | 13.521 | 4.3079 | 0.7238 | 32177.28 | 716.8466 | 120611.9 | 39137.8242 |
| | 3 | 26.74 | 4.27 | 0 | 0 | 17.701 | 2.308 | 102.104 | 9.346 | 4.5363 | 0.765 | 32413.09 | 581.7217 | 161630.7 | 37035.2625 |
| | 4 | 29.87 | 4.493 | 0 | 0 | 14.258 | 2.518 | 87.075 | 12.468 | 4.397 | 0.7395 | 32032.87 | 750.0289 | 118041.1 | 36646.8146 |
| | 5 | 35.19 | 5.142 | 0.006 | 0.077 | 9.059 | 2.676 | 58.733 | 16.013 | 4.0844 | 0.6768 | 31237.8 | 919.8058 | 65527.05 | 32220.9459 |
| | 6 | 32.81 | 4.649 | 0 | 0 | 12.138 | 2.64 | 76.712 | 14.884 | 4.417 | 0.748 | 32117.32 | 801.4451 | 90607.97 | 32073.7435 |
| | 7 | 36.364 | 4.946 | 0.002 | 0.045 | 8.258 | 2.656 | 54.108 | 16.584 | 4.2589 | 0.7155 | 31191.78 | 946.228 | 56523.23 | 28417.2914 |
| | 8 | **44.082** | 4.463 | **0.064** | 0.303 | **0.976** | 1.184 | **6.054** | 8.46 | 4.0538 | 0.6846 | 28331.11 | 1073.477 | **5418.824** | 8430.7514 |
| | 9 | 5.278 | 1.58 | 0 | 0 | 18.821 | 1.434 | 85.456 | 4.305 | **6.7045** | 1.1244 | 31253.08 | 456.2674 | 286050.1 | 35533.4621 |
| | 10 | 1 | **0** | 0 | 0 | 20.393 | **0.106** | **5.148** | **3.944** | 2.6425 | 0.3924 | **28024.24** | **44.3314** | 499880.3 | **1086.7815** |
| | 11 | 8.528 | 2.42 | 0 | 0 | 23.357 | 3.072 | 109.155 | 10.135 | 0.013 | **0.022** | 32648.83 | 926.4653 | 272009.9 | 62082.4709 |
| 100 | 0 | 27.056 | 4.507 | 0 | 0 | 16.977 | 0.907 | 113.305 | 3.756 | 17.8248 | 2.9748 | 65051.87 | 667.284 | 558643.9 | 50252.7832 |
| | 1 | 28.476 | 4.388 | 0 | 0 | 15.963 | 1.018 | 111.674 | 4.176 | 17.1294 | 2.739 | 64873.31 | 824.3303 | 509758.2 | 53979.6751 |
| | 2 | 31.5 | 5.173 | 0 | 0 | 14.406 | 1.139 | 107.424 | 5.392 | 16.6702 | 2.6083 | 64582.33 | 1120.178 | 438233.4 | 60357.1999 |
| | 3 | 30.116 | 4.689 | 0 | 0 | 15.331 | 0.948 | 110.571 | 4.615 | 17.569 | 2.8203 | 64877.63 | 892.6177 | 478286.6 | 49968.1420 |
| | 4 | 32.442 | 4.55 | 0 | 0 | 13.563 | 1.04 | 104.043 | 5.577 | 16.9932 | 2.6475 | 64347.25 | 1159.757 | 402642.2 | 55190.6790 |
| | 5 | 37.524 | 5.4 | 0 | 0 | 10.642 | 1.067 | 87.882 | 6.868 | 15.7426 | 2.3916 | 62982.56 | 1581.812 | 302505.7 | 58114.0986 |
| | 6 | 37.188 | 4.95 | 0.002 | 0.045 | 11.909 | 1.066 | 96.473 | 6.802 | 17.1704 | 2.702 | 64295.33 | 1245.893 | 334714.5 | 50535.6466 |
| | 7 | 40.262 | 5.529 | 0.002 | 0.045 | 9.49 | 1.126 | 80.679 | 8.005 | 16.5428 | 2.4717 | 62883.59 | 1635.266 | 260267.7 | 55235.1007 |
| | 8 | **57.302** | 6.849 | **0.152** | 0.435 | **2.329** | 0.992 | 21.916 | 9.291 | 15.8048 | 2.3262 | **55830.02** | 2048.334 | **57761.36** | 42030.1210 |
| | 9 | 6.848 | 1.843 | 0 | 0 | 13.321 | 0.697 | 85.525 | 3.172 | **25.8886** | 3.993 | 62458.84 | 631.2151 | 573135.1 | 48380.8306 |
| | 10 | 1 | **0** | 0 | 0 | 14.418 | **0.054** | **5.869** | **2.966** | 10.1184 | 1.161 | 56046.99 | **62.3825** | 999720.8 | **1646.6140** |
| | 11 | 10.444 | 2.663 | 0 | 0 | 17.906 | 1.378 | 109.495 | 7.892 | 0.0476 | **0.05** | 65316.82 | 1320.111 | 621792.5 | 85797.1254 |
| 150 | 0 | 29.234 | 4.409 | 0 | 0 | 14.51 | 0.548 | 111.228 | 3.186 | 39.0909 | 5.8462 | 97500.52 | 813.8773 | 924285.2 | 56397.7725 |
| | 1 | 30.044 | 4.752 | 0.004 | 0.063 | 13.987 | 0.624 | 111.209 | 3.446 | 37.5183 | 5.4441 | 97367.75 | 1050.166 | 873603.5 | 62586.7269 |
| | 2 | 32.602 | 4.976 | 0 | 0 | 12.975 | 0.721 | 109.937 | 3.887 | 36.3396 | 5.1288 | 97031.77 | 1314.433 | 781236.9 | 72066.7149 |
| | 3 | 32.566 | 4.636 | 0 | 0 | 13.238 | 0.588 | 110.849 | 3.482 | 38.4948 | 5.5112 | 97373.33 | 1186.611 | 801509.8 | 59333.8052 |
| | 4 | 34.552 | 4.915 | 0 | 0 | 12.02 | 0.672 | 106.822 | 4.391 | 37.3443 | 5.448 | 96646.12 | 1497.259 | 702948.2 | 67389.4702 |
| | 5 | 38.842 | 5.225 | 0 | 0 | 9.865 | 0.657 | 94.662 | 5.112 | 34.557 | 4.8437 | 94766.85 | 2117.23 | 561339.4 | 76923.5292 |
| | 6 | 39.38 | 4.876 | 0.002 | 0.045 | 10.63 | 0.684 | 100.761 | 5.231 | 37.7004 | 5.5096 | 96480.24 | 1760.161 | 593359.7 | 66298.1134 |
| | 7 | 43.326 | 5.156 | 0.002 | 0.045 | 8.652 | 0.66 | 86.662 | 5.569 | 36.3051 | 5.1832 | 94508.79 | 2363.881 | 476971 | 77652.8961 |
| | 8 | **62.466** | 7.329 | **0.336** | 0.713 | **2.584** | 0.702 | 29.27 | 7.644 | 34.4403 | 4.8116 | **83208.2** | 2666.568 | **142370.8** | 69677.5283 |
| | 9 | 7.816 | 1.896 | 0 | 0 | 10.879 | 0.497 | 85.598 | 2.492 | **56.5194** | 8.1362 | 93759.12 | 788.4589 | 857348.1 | 63773.8475 |
| | 10 | 1 | **0** | 0 | 0 | 11.777 | **0.036** | **5.749** | **2.406** | 22.2228 | 2.7068 | 84077.84 | **77.1704** | 1499801 | **1397.3776** |
| | 11 | 11.27 | 2.609 | 0 | 0 | 15.074 | 0.927 | 107.625 | 7.159 | 0.1059 | **0.0684** | 97772.24 | 1688.187 | 994780.5 | 101431.9720 |

**TABLE 5.** Result for large size problem.

| $N_{sol}$ | Alg | Avg($N_n$) | Std($N_n$) | Avg($N_p$) | Std($N_p$) | Avg(GD) | Std(GD) | Avg(SP) | Std(SP) | Avg(T) | Std(T) | Avg($F_e$) | Std($F_e$) | Avg($F_m$) | Std($F_m$) |
|---|---|---|---|---|---|---|---|---|---|---|---|---|---|---|---|
| 50 | 0 | 27.174 | 4.155 | 0 | 0 | 748.62 | 96.656 | 4480.39 | 365.757 | 8.4167 | 1.5329 | 98628.83 | 771.7182 | 143665.5 | 35722.9236 |
| | 1 | 29.798 | 4.686 | 0 | 0 | 639.303 | 109.316 | 4019.28 | 526.113 | 8.0533 | 1.343 | 98279.76 | 960.7157 | 106329.4 | 34186.9693 |
| | 2 | 34.79 | 5.074 | 0 | 0 | 478.211 | 136.159 | 3161.886 | 809.014 | 7.7758 | 1.2132 | 97705.9 | 1259.995 | 62884.07 | 30649.8855 |
| | 3 | 30.616 | 4.385 | 0 | 0 | 633.698 | 108.007 | 3994.477 | 524.318 | 8.218 | 1.3142 | 98354.38 | 893.9128 | 104469.4 | 33485.2153 |
| | 4 | 34.058 | 4.867 | 0 | 0 | 484.791 | 118.408 | 3210.835 | 695.26 | 7.9161 | 1.2229 | 97382.84 | 1207.66 | 63353.57 | 27636.0545 |
| | 5 | 42.104 | 5.073 | 0 | 0 | 237.875 | 158.78 | 1641.271 | 1075.713 | 7.293 | 1.092 | 95492.67 | 1530.155 | 21440.25 | 19011.2797 |
| | 6 | 37.198 | 4.822 | 0 | 0 | 408.566 | 128.757 | 2754.903 | 808.218 | 7.904 | 1.2089 | 97480.85 | 1267.524 | 46923.7 | 24637.2246 |
| | 7 | 42.954 | 4.724 | 0.002 | 0.045 | 205.495 | 154.737 | 1424.205 | 1060.277 | 7.6309 | 1.108 | 95610.53 | 1573.731 | 17520.32 | 16311.2192 |
| | 8 | **45.918** | 2.998 | **0.006** | 0.1 | **6.73** | 36.201 | **45.981** | 255.548 | 6.8534 | 1.0066 | **89370.73** | 2219.432 | **1315.029** | 2029.9054 |
| | 9 | 6.328 | 1.726 | 0 | 0 | 904.797 | 76.097 | 4909.066 | **144.96** | **12.4131** | 1.93 | 97663.75 | 709.6203 | 207628.7 | 34209.3182 |
| | 10 | 4.948 | **1.556** | 0 | 0 | 1152.116 | 58.678 | 4687.853 | 261.903 | 4.7327 | 0.6193 | 98649.82 | **634.5005** | 334246.6 | 33687.3944 |
| | 11 | 10.368 | 2.799 | 0 | 0 | 856.652 | 145.453 | 4711.009 | 392.009 | 0.0245 | **0.0245** | 98891.73 | 1471.68 | 190118.4 | 61317.1989 |
| 100 | 0 | 30.448 | 4.494 | 0 | 0 | 657.704 | 35.537 | 4947.358 | **69.806** | 33.2646 | 5.1487 | 197248.5 | 1061.073 | 436547.4 | 46564.5095 |
| | 1 | 32.06 | 4.894 | 0 | 0 | 613.194 | 42.945 | 4832.571 | 135.141 | 31.8998 | 4.8814 | 196881.6 | 1430.933 | 380499.4 | 52806.8613 |
| | 2 | 36.274 | 5.131 | 0 | 0 | 539.157 | 47.963 | 4531.276 | 241.08 | 30.8796 | 4.6327 | 196021.5 | 1827.626 | 295533.3 | 51535.3657 |
| | 3 | 34.158 | 4.699 | 0 | 0 | 586.502 | 47.804 | 4742.246 | 167.804 | 32.2676 | 4.963 | 196844.5 | 1456.553 | 348421.8 | 49963.8885 |
| | 4 | 37.156 | 5.092 | 0 | 0 | 511.663 | 48.763 | 4388.164 | 277.232 | 31.3592 | 4.5905 | 195426.1 | 2058.278 | 266699.6 | 49262.8425 |
| | 5 | 44.196 | 5.801 | 0 | 0 | 400.691 | 56.641 | 3663.083 | 419.515 | 28.8562 | 4.0773 | 192100.9 | 2664.295 | 166154.7 | 45483.4355 |
| | 6 | 41.918 | 5.683 | 0 | 0 | 453.726 | 49.978 | 4037.566 | 329.065 | 31.453 | 4.5322 | 195306.7 | 2181.89 | 210798.2 | 45526.3109 |
| | 7 | 47.33 | 5.836 | 0 | 0 | 363.759 | 57.152 | 3510.288 | 449.962 | 30.117 | 4.1393 | 191714.2 | 2919.089 | 137930.9 | 41874.8504 |
| | 8 | **72.528** | 7.324 | **0.116** | 0.468 | **38.469** | 58.147 | **382.143** | 578.595 | 27.616 | 3.7534 | **176917.2** | 4417.363 | **6889.327** | 8665.4426 |
| | 9 | 7.948 | 2.065 | 0 | 0 | 641.551 | 38.106 | 4910.514 | 90.919 | **49.2814** | 7.0987 | 195334.2 | 995.1512 | 416075.8 | 48864.6986 |
| | 10 | 6.508 | **1.72** | 0 | 0 | 813.33 | **28.137** | 4705.793 | 166.873 | 19.1044 | 2.2699 | 197323 | **843.5352** | 665380.8 | 45753.2918 |
| | 11 | 12.482 | 2.856 | 0 | 0 | 686.503 | 64.443 | 4926.057 | 119.871 | 0.0794 | **0.041** | 197624.3 | 2143.287 | 478296.4 | 82782.5684 |
| 150 | 0 | 31.892 | 4.499 | 0 | 0 | 571.334 | 23.553 | 4986.075 | **23.806** | 74.9205 | 11.0373 | 295934.7 | 1364.945 | 739913.7 | 60689.2869 |
| | 1 | 33.948 | 4.799 | 0 | 0 | 541.511 | 25.246 | 4952.636 | 54.778 | 71.7156 | 10.4578 | 295532.6 | 1663.517 | 665347.3 | 61362.5540 |
| | 2 | 37.058 | 5.306 | 0 | 0 | 497.394 | 31.575 | 4816.909 | 128.206 | 69.2553 | 9.7601 | 294401.4 | 2369.671 | 562929.2 | 70555.4639 |
| | 3 | 36.176 | 5.035 | 0 | 0 | 515.033 | 26.975 | 4884.23 | 88.475 | 73.4922 | 10.7752 | 295225.5 | 1901.874 | 602559.3 | 62546.5565 |
| | 4 | 38.654 | 5.274 | 0 | 0 | 461.168 | 33.508 | 4648.141 | 179.439 | 70.7016 | 10.1622 | 293575.6 | 2662.045 | 485020.5 | 69694.8734 |
| | 5 | 46.074 | 6.107 | 0.002 | 0.045 | 375.99 | 38.045 | 4077.306 | 301.451 | 65.0655 | 9.094 | 289048.3 | 3537.873 | 325141.9 | 64553.0441 |
| | 6 | 44.548 | 5.498 | 0 | 0 | 407.532 | 32.447 | 4317.863 | 231.212 | 70.9386 | 10.2171 | 292870.2 | 2835.374 | 379900.7 | 59375.2787 |
| | 7 | 49.592 | 5.643 | 0.004 | 0.063 | 339.382 | 35.921 | 3774.411 | 317.912 | 68.0226 | 9.4273 | 288517.2 | 3575.387 | 265771.2 | 54642.6111 |
| | 8 | **77.03** | 7.705 | **0.362** | 1.141 | **75.415** | 53.721 | **915.721** | 647.916 | 62.04 | 8.258 | **263672.3** | 6382.976 | **22508.07** | 20508.0012 |
| | 9 | 8.758 | 2.175 | 0 | 0 | 522.934 | 26.468 | 4906.145 | 78.318 | **110.3736** | 14.8236 | 293040.9 | 1222.526 | 621413.8 | 62184.5910 |
| | 10 | 7.398 | **2.044** | 0 | 0 | 666.568 | **18.525** | 4689.758 | 137.576 | 43.2861 | 4.9709 | 295948.8 | **1105.613** | 1005113 | 55590.2798 |
| | 11 | 13.36 | 3.007 | 0 | 0 | 590.492 | 41.685 | 4939.701 | 88.364 | 0.18 | **0.0623** | 296189.8 | 2758.174 | 792818.2 | 110125.8133 |

## 6. Conclusion

This study sheds light on a nascent bi-objective time-constrained task scheduling problem focusing on energy consumption and service quality in terms of the makespan to find non-dominated solutions for the purpose of ameliorating the service quality and addressing environmental issues of cloud computing services. In response to this bi-objective problem, we proposed a new bi-objective simplified swarm optimization (BSSO) algorithm.

To ensure better solution quality, the BSSO algorithm integrates the crowding distance, a hybrid elite selection, and a new stepwise update mechanism. From the experiments conducted on three different sized problems [17], regardless of the parameter setting, each of the proposed BSSO outperformed the MOPSO [13], MOSSO [17], and NSGA-II [19], in convergency, diversity, the number of obtained temporary nondominated solutions, and the number of obtained real nondominated solutions. Among nine different parameter settings, we concluded that the BSSO algorithm with $c_p = c_w = 0.5$ is the best one. The results prove that the proposed BSSO can successfully achieve the aim of this work.

## Acknowledgments

This research was supported in part by the National Science Council of Taiwan, R.O.C. under grant MOST 101-2221-E-007-079-MY3 and MOST 102-2221-E-007-086-MY3.